\documentclass[twocolumn,twosided]{IEEEtran}

\usepackage{url}
\usepackage{graphicx}
\usepackage{amsmath}
\usepackage{multirow}
\usepackage{amssymb}
\usepackage{wrapfig}
\usepackage{tikz}
\usetikzlibrary{shapes,arrows}
\usepackage{float}
\usepackage{sidecap}
\usepackage{placeins}
\usepackage[shortlabels]{enumitem}
\begin{document}

\title{Cross-Spectrum Measurement Statistics}

\author{Antoine Baudiquez, \'{E}ric Lantz, Enrico Rubiola, Fran\c{c}ois Vernotte
\thanks{A. Baudiquez and E. Rubiola are with FEMTO-ST, Department of Time and Frequency, UMR 6174, Universit\'{e} Bourgogne Franche-Comt\'{e}, France}
\thanks{E. Lantz is with FEMTO-ST, D\'{e}partement d'Optique P.M. Duffieux, UMR 6174 CNRS, Universit\'{e} Bourgogne Franche-Comt\'{e}, France}
\thanks{E. Rubiola is with Physics Metrology Division, Istituto Nazionale di Ricerca MetrologicaINRiM, Torino, Italy}
\thanks{F. Vernotte is with FEMTO-ST, Department  of  Time  and  Frequency, Observatory  THETA, UMR 6174 CNRS, Universit\'{e} Bourgogne Franche-Comt\'{e}, France}}

\markboth{IEEE TRANSACTION ON ULTRASONICS, FERROELECTRICS, AND FREQUENCY CONTROL}
{Shell \MakeLowercase{\textit{et al.}}: Bare Demo of IEEEtran.cls for IEEE Journals}
\maketitle

\begin{abstract}
%
%
%
The cross-spectrum method consists in measuring a signal $c(t)$ simultaneously with two independent instruments. Each of these instruments contributes to the global noise by its intrinsec (white) noise, whereas the signal $c(t)$ that we want to characterize could be a (red) noise.

We first define the real part of the cross-spectrum as a relevant estimator. Then, we characterize the probability density function (PDF) of this estimator knowing the noise level (direct problem) as a Variance-Gamma (VG) distribution. Next, we solve the ``inverse problem'' thanks to Bayes' theorem to obtain an upper limit of the noise level knowing the estimate. Checked by massive Monte Carlo simulations, VG proves to be perfectly reliable to any number of degrees of freedom (dof).

Finally we compare this method with an other method using the Karhunen-Lo\`{e}ve transfrom (KLT). We find an upper limit of the signal level slightly different as the one of VG since KLT better takes into account the available informations.
\end{abstract}

\begin{IEEEkeywords}
Bayesian statistics, confidence interval, cross-spectrum, Karhunen-Lo\`{e}ve transform, probability density function.
\end{IEEEkeywords}

\IEEEpeerreviewmaketitle

\section{Introduction}
\IEEEPARstart{T}{he} measurement of power spectra is a classical problem, ubiquitous in numerous branches of physics, as explained below.  Power spectra are efficiently measured using Fourier transform methods with digitized data.  Relevant bibliography is now found in classic books \cite{blackman1959,jenkins1968,brigham1988,percival1993}.

We are interested in the measurement of weak statistical phenomena, which challenge the instruments and the mathematical tools, using the cross-spectrum method. This method consists of the simultaneous measurement of the signal with two separate and independent instruments \cite{rubiola2010}.  The other approach, consisting on the observation of the spectral contrast in a chopped signal, broadly equivalent to the Dicke radiometer \cite{dicke1946}, is not considered here. Regarding the duration of the data record used to evaluate the Fast Fourier Transform (FFT), two asymptotic cases arise.

The first case is that of the measurement of fast phenomena, where a large number of records denoted $m$ is possible in a reasonable observation time.  At large $m$ the central limit theorem rules and the background noise can be rejected by a factor approximately equal to $1/\sqrt{m}$ depending on the estimator.
Numerous examples are found in the measurement of noise in semiconductors \cite{sampietro1999}, phase noise in oscillators and components \cite{hati2016,gruson2017,cardenas2017,feldhaus2016}, frequency fluctuations and relative intensity noise in lasers \cite{fortier2012,rubiola2005-arXiv}, electromigration in thin films \cite{Verbruggen1989}, etc.  Restricting to one bin of the Fourier transform, the power spectral density integrated over a suitable frequency range is used in radiometry \cite{allred1962,nanzer2008}, Johnson thermometry \cite{white1996} and other applications.

The second case is that of slow phenomena, where the fluctuations are very long term or non ergodic. On one hand the background noise is still rejected as before but with a very low $m$ which can actually be equal to one. On the other hand, the central limit theorem does not apply and the statistical uncertainty are not trivial.
This case is of great interest in radio astronomy, where the observations are limited by the available resources and take long time. As instance millisecond pulsars (MSP) can be used as very stable clocks at astronomical distances \cite{verbiest2009}. The radio pulses time of arrival (TOA) of MSP are affected by numerous physical process, one on them are gravitational-wave (GW) perturbations \cite{detweiler1979,hellings1983}. Red noise originate from GW perturbations in the signal path common to the radio-telescopes can be detected \cite{taylor2016,perrodin2018}.
Like the analysis of the signals provided by the LIGO/VIRGO interferometers which use cross correlation methods \cite{drasco2003,cornish2013}, the LEAP experiment (i.e. Large European Array for Pulsars) \cite{bassa2015} could use such methods in order to access lower frequencies and observe imperceptible phenomena such as early phases well before the coalescence of black holes or GW of cosmological origin (for example cosmic strings, inflation, primordial black holes).

This article is intended to put an upper limit on the uncertainty of the cross-spectrum estimate. Gravitational-waves have not been yet discovered in the TOA of millisecond pulars. Thanks to the very long line of sight between the pulsars and us (several thousand light-years), we could access very low frequencies, inaccessible to LIGO/VIRGO, thus revealing much slower astrophysical phenomena. It is therefore important to develop statistical tools to improve measurement sensitivity. In this respect, we propose in Section \ref{sec1} to state the cross-spectrum problem to define a proper estimate. Based on the principle that the experiment is repeated $m$ times, it is important to note that the estimation of the measurement uncertainty is A-Type as defined by the VIM \cite{VIM2004}. Then in Section \ref{s:direct} we define the probability density function (i.e. ``direct problem'') of the cross-spectrum estimate which is used in Section \ref{sec3} to compute an upper limit by using a bayesian inference approach (i.e. ``inverse problem''). The result obtained are compared to an another method using the Karhunen-Lo\`{e}ve transform developed in \cite{lantz2019} and the conclusions are presented in Section \ref{sec4}.

\begin{figure}
\centerline{\includegraphics[width=\columnwidth]{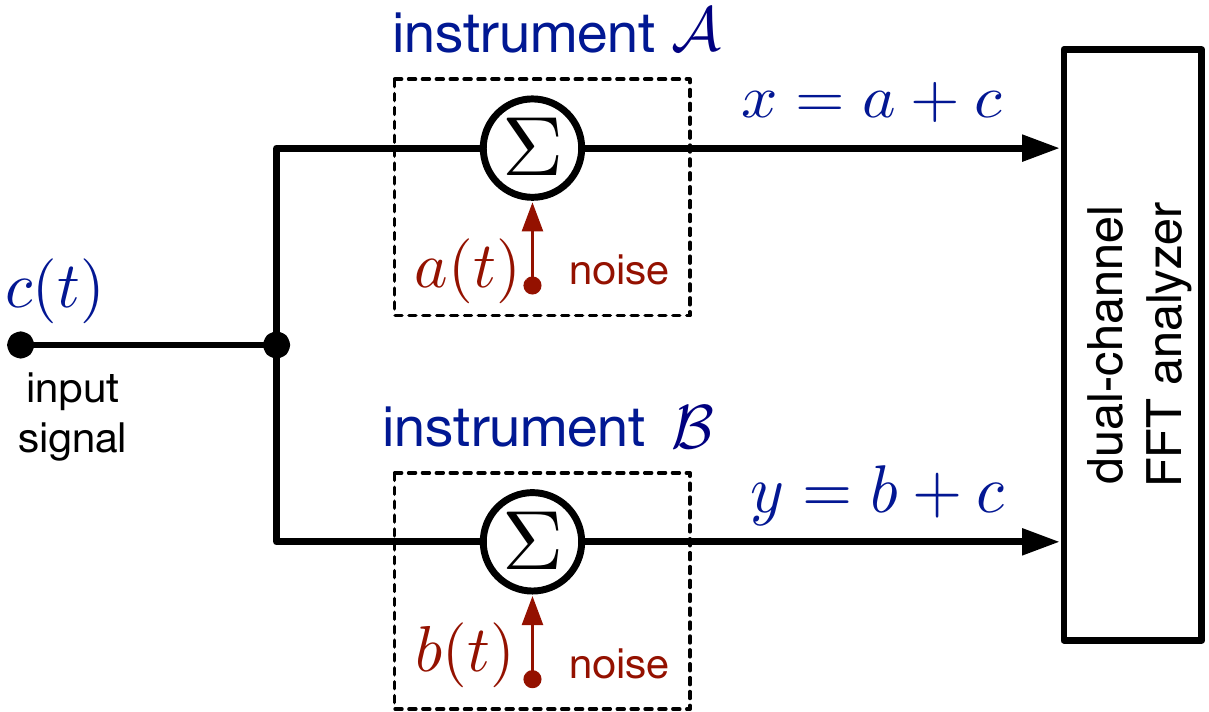}}
\caption{Basics of the cross-spectrum method.}
\label{fig1}
\end{figure}

\section{Statement of the Problem\label{sec1}}

\subsection{Cross-spectrum method}

Let us consider 3 statistically independent signals: $a(t)$, $b(t)$ and $c(t)$ as shown in Fig. \ref{fig1}. On one side the two first $a(t)$ and $b(t)$ are respectively the instrument noise of $\mathcal{A}$ and $\mathcal{B}$. On the other side $c(t)$ is an input signal which we want to characterize. In the case of pulsar measurement, this input signal is also a noise, but in general a red noise. The output of each channel is

\begin{equation}
    \begin{aligned}
        x(t)&=a(t)+c(t)\\
        y(t)&=b(t)+c(t).
    \end{aligned}
    \label{eq1}
\end{equation}

Applying the Fourier Transform (FT) on each channel gives

\begin{equation}
    \begin{aligned}
    X(f)&=A(f)+C(f)\\
    Y(f)&=B(f)+C(f)
    \end{aligned}
    \label{eq2}
\end{equation}
\noindent
where $f$ is the frequency, $X(f)$, $Y(f)$, $A(f)$, $B(f)$ and $C(f)$ stand respectively for the Fourier Transform (FT) of $x(t)$, $y(t)$, $a(t)$, $b(t)$ and $c(t)$. Our interest is carried out on the power spectral density (PSD) rather than the spectrum. The cross-spectrum is defined as

\begin{equation}
    S_{yx}(f)=\frac{1}{T}Y(f)X^*(f),
    \label{eq3}
\end{equation}
\noindent
where the cross-spectrum is actually a cross-PSD and the factor $\frac{1}{T}$ is the measurement time which is necessary to have the dimension of a power per frequency. The $^*$ denotes the complex conjugate of the quantity placed before it. Experimentaly averaging over $m$ spectra realizations leads to the following relation

\begin{equation}
\left\langle S_{yx}\right\rangle_m=\frac{1}{T}\left\langle Y(f)X^*(f) \right\rangle_m.
\label{eq4}
\end{equation}

\subsection{Cross Power Spectral Density}

Averaging on a large number of observations, the mathematics is made simple by the central limit theorem, by which all the probability density functions (PDF) become Gaussian. More interesting for us is the case of a small number of realizations, each of which taking long observation time-up to several years in the case of the millisecond pulsars.\\
The random variables (rv) $a(t)$, $b(t)$ and $c(t)$ follow a centered Normal distribution whatever the kind of noise. Even red noise (e.g. random walk) follows a Normal distribution not on the time average but regarding its ensemble average over the probability space which means it is a non ergodic process. Moreover a stochastic process with zero-mean Gaussian distribution has a FT which is also a random process with centered Gaussian distribution.\\
The rv $A(f)$, $B(f)$ and $C(f)$ can then be decomposed into a real and imaginary part

\begin{equation}
    \begin{aligned}
        A(f)&=A'(f)+iA''(f)\\
        B(f)&=B'(f)+iB''(f)\\
        C(f)&=C'(f)+iC''(f).
    \end{aligned}
    \label{eq5}
\end{equation}

The real and imaginary part are statistically independent rv with equal variance following a zero-mean Gaussian distribution. In the following, we will omit the frequency dependence $f$ because we are working only in the Fourier domain. Let us now expand Eq. \ref{eq4},

\begin{equation}
    \begin{aligned}
        \left\langle S_{yx}\right\rangle_m&=\frac{1}{T}\left\langle YX^* \right\rangle_m.\\
        &=\frac{1}{T}\left[\left\langle A'B'+B'C'+C'A'+C'^{\,2}\right\rangle_m\right.\\
        &\hspace{4mm} +\left\langle A''B''+B''C''+C''A''+C''^{\,2}\right\rangle_m\\
        &\hspace{4mm} +i\left\langle A'B''+B''C'+C''A'\right\rangle_m\\
        &\hspace{4mm} -i\left.\left\langle A''B'+B'C''+C'A''\right\rangle_m\right].
    \end{aligned}
    \label{eq6}
\end{equation}

The terms in the imaginary part have a zero expectation, while the expectation in the real part is proportional to the PSD of the signal, i.e. what we are looking to characterize. As a consequence, the next sections focus solely on the real part $\mathcal{R}\left\{S_{yx}\right\}$,

\begin{equation}
    \left\langle\mathcal{R}\left\{S_{yx}\right\}\right\rangle_m=\frac{1}{T}\left\langle(A^k+C^k)(B^k+C^k)\right\rangle_\nu
    \label{eq7}
\end{equation}
\noindent
where $\nu=2m$ the number of degree of freedom (dof). The superscript $k$ means real or imaginary part because they are independent rv.

\subsection{Definition of the problem}
\subsubsection{Measurements, and estimates}

In the following, in order to simplify the notation, we will omit the superscript $k$. Thereby the real and imaginary part will be treated as 2 dof. Moreover to simplifiy the notations, we will omit the factor $\frac{1}{T}$ which does not affect the PDF.\\
We refer the cross-spectrum measurement for a given frequency to

\begin{equation}
    Z=(A+C)(B+C)
    \label{eq8}
\end{equation}
\noindent
where all $A$, $B$, $C$ are rv which are independent, centered and Normal. In the following, we will assume that $A$, $B$, $C$ have only 1 dof, their real or their imaginary part, and that $Z$ does not come from the average of different spectra. A generalization of this problem to 2 dof (real and imaginary parts) and then 2$m$ dof (average of $m$ spectra) will be given.\\
The \textbf{estimates} are denoted $\hat{\sigma}_A^2$, $\hat{\sigma}_B^2$, $\hat{\sigma}_C^2$ which respectively correpond to the variance of $A$, $B$, $C$.

\subsubsection{Direct and Inverse Problem}

In order to assess the uncertainty over the estimator $\sigma_C^2$, called the \textit{signal} level, we will have to distinguish to main issues:
\begin{itemize}
    \item The \textbf{direct problem} consists in calculating the statistics of the cross-spectrum measurement $Z$, knowing the model parameters $\sigma_A^2$, $\sigma_B^2$, $\sigma_C^2$.
    \item The \textbf{inverse problem}, conversely consists in calculating a confidence interval over the model parameter $\sigma_C^2$, from the estimates $\sigma_A^2$, $\sigma_B^2$ and the cross-spectrum measurement $Z$.
\end{itemize}

\section{Direct Problem\label{s:direct}}
\subsection{Vector formalization of the problem}

We will reuse here the formalism we developed in \cite{vernotte2019}, i.e. a vector space of normal laws. Since we have 3 Normal rv, we are in a vector space of 3 dimensions that we will denote $\mathbb{LG}^3$ and which has the basis $(\vec{E}_A,\vec{E}_B,\vec{E}_C)$ defined as
$$
\left\{\begin{array}{l}
\vec{E}_A=\textrm{LG}_A(0,1)\\
\vec{E}_B=\textrm{LG}_B(0,1)\\
\vec{E}_C=\textrm{LG}_C(0,1)
\end{array}\right.
$$
where LG$(0,1)$ stands for a Laplace-Gauss (or Normal) rv. with zero-mean (centered) and unity standard deviation ($\sigma=1$). We assume that LG$_A(0,1)$, LG$_B(0,1)$, LG$_C(0,1)$ are independent.
Any vector $\vec{U}$ may be written as
$$
\vec{U}=\left(\begin{array}{c}
u_A\\
u_B\\
u_C
\end{array}\right)=u_A\vec{E}_A+u_B\vec{E}_B+u_C\vec{E}_C
$$
where $u_A,u_B,u_C$ are 3 constant scalars since all the random part is carried by the basis vectors.
The scalar product between 2 vectors $\vec{U}$ and $\vec{V}$ is defined as:
\begin{eqnarray}
\vec{U}\cdot \vec{V}&=&\left(u_A\vec{E}_A+u_B\vec{E}_B+u_C\vec{E}_C\right)\cdot\nonumber\\
&&\left(v_A\vec{E}_A+v_B\vec{E}_B+v_C\vec{E}_C\right)\nonumber\\
&=&u_Av_A\vec{E}_A\cdot\vec{E}_A+u_Bv_B\vec{E}_B\cdot\vec{E}_B+u_Cv_C\vec{E}_C\cdot\vec{E}_C\nonumber\\
&&+(u_Av_B+u_Bv_A)\vec{E}_A\cdot\vec{E}_B\nonumber\\
&&+(u_Bv_c+u_Cv_B)\vec{E}_B\cdot\vec{E}_C\nonumber\\
&&+(u_Cv_A+u_Av_C)\vec{E}_C\cdot\vec{E}_A.\nonumber
\end{eqnarray}
From this definition of the scalar product, we see that
$$
\left\{\begin{array}{l}
||\vec{E}_A||^2=\vec{E}_A\cdot\vec{E}_A=\textrm{LG}_A\cdot\textrm{LG}_A=\chi_A^2\\
||\vec{E}_B||^2=\vec{E}_B\cdot\vec{E}_B=\textrm{LG}_B\cdot\textrm{LG}_B=\chi_B^2\\
||\vec{E}_C||^2=\vec{E}_C\cdot\vec{E}_C=\textrm{LG}_C\cdot\textrm{LG}_C=\chi_C^2
\end{array}\right.
$$
$$
\left\{\begin{array}{l}
\vec{E}_A\cdot\vec{E}_B=\textrm{LG}_A\cdot\textrm{LG}_B=\textrm{V}\Gamma_{AB}\\
\vec{E}_B\cdot\vec{E}_C=\textrm{LG}_B\cdot\textrm{LG}_C=\textrm{V}\Gamma_{BC}\\
\vec{E}_C\cdot\vec{E}_A=\textrm{LG}_C\cdot\textrm{LG}_A=\textrm{V}\Gamma_{CA}
\end{array}\right.
$$
where $\chi_{A,B,C}^2$ are 3 independent $\chi^2$ rv with 1 dof and V$\Gamma_{AB,BC,CA}$ are 3 variance-Gamma (V$\Gamma$) rv \cite{vernotte2019,haas2009}.

On the other hand, if we consider the mathematical expectation of these expressions, we obtain
$$
\mathbb{E}\left[\vec{E}_P\cdot\vec{E}_Q\right]=\delta_{P,Q} \quad \textrm{with} \quad P, Q \in \left\{A,B,C\right\}
$$
where the $\mathbb{E}[\cdot]$ stands for the mathematical expectation of the quantity within the brackets and $\delta_{P,Q}$ is the Kronecker delta.
We see that we obtain the classical scalar product by using the mathematical expectation:
$$
\mathbb{E}\left[\vec{U}\cdot \vec{V}\right]=u_Av_A+u_Bv_B+u_Cv_C.
$$
Therefore, we will define that 2 vectors $\vec{U}$ and $\vec{V}$ are orthogonal if $\mathbb{E}\left[\vec{U}\cdot\vec{V}\right]=0$.

\subsection{From a normal random variable product to a chi-squared rv difference}
Following this formalism, Eq. (\ref{eq8}) may be rewritten as
\begin{equation}
    \begin{aligned}
Z&=\left(\vec{A}+\vec{C}\right)\cdot\left(\vec{B}+\vec{C}\right)=\left(\begin{array}{c}
a\\
0\\
c\end{array}\right)\cdot\left(\begin{array}{c}
0\\
b\\
c\end{array}\right)\\
&=ab\textrm{V}\Gamma_{AB}+ac\textrm{V}\Gamma_{AC}+bc\textrm{V}\Gamma_{BC}+c^2\chi_C^2\label{eq:rvX}
\end{aligned}
\end{equation}
where $a,b,c$ are respectively the standard deviations of the rv $A,B,C$. As a consequence, $\mathbb{E}[X]=c^2$. In the following, we will use the noise variances $\sigma_A^2=a^2, \sigma_B^2=b^2$ and the signal variance $\sigma_C^2=c^2$.

As demonstrated in \cite{sorin1968}, a product of independent normal rv may be expressed as a difference of $\chi^2$ rv. For this purpose, although we know that $(A+C)$ and $(B+C)$ are not independent, we introduce the rv $V_1=(A+B)/2+C$ and $V_2=(A-B)/2$ in such a way that $A+C=V_1+V_2$, $B+C=V_1-V_2$ and therefore $(A+C)(B+C)=V_1^2-V_2^2$. In this vectorial formalism:
$$
\vec{V}_1=\left(\begin{array}{c}
a/2\\
b/2\\
c
\end{array}\right), \quad \textrm{and} \quad \vec{V}_2=\left(\begin{array}{c}
a/2\\
-b/2\\
0
\end{array}\right).
$$
Therefore, $(\vec{V}_1,\vec{V}_2)$ is the basis of the 2-dimensional subspace of $\mathbb{LG}^3$ in which lies our whole problem.
Since the squared modulus of $\vec{V}_1, \vec{V}_2$ are:
$$
\left\{\begin{array}{lcl}
||\vec{V}_1||^2&=&\displaystyle\frac{a^2}{4}\chi_A^2+\frac{b^2}{4}\chi_B^2+c^2\chi_C^2\nonumber\\
&&+\frac{ab}{2}\textrm{V}\Gamma_{AB}+ac\textrm{V}\Gamma_{AC}+bc\textrm{V}\Gamma_{BC}\nonumber\\
||\vec{V}_2||^2&=&\displaystyle\frac{a^2}{4}\chi_A^2+\frac{b^2}{4}\chi_B^2-\frac{ab}{2}\textrm{V}\Gamma_{AB},
\end{array}\right.
$$
their difference is consistent with Eq. (\ref{eq:rvX}) and then $Z=(\vec{A}+\vec{C})\cdot(\vec{B}+\vec{C})=||\vec{V}_1||^2-||\vec{V}_2||^2$.
Moreover, we can calculate the mathematical expectations of these squared modulus:
\begin{equation}
\begin{aligned}
v_1^2&=\mathbb{E}\left[||\vec{V}_1||^2\right]=\frac{a^2+b^2}{4}+c^2\\
v_2^2&=\mathbb{E}\left[||\vec{V}_2||^2\right]=\frac{a^2+b^2}{4}.\label{eq:v12v22}
\end{aligned}
\end{equation}
On the other hand, since
\begin{equation}
\mathbb{E}\left[\vec{V}_1\cdot\vec{V}_2\right]=\frac{a^2-b^2}{4}\label{eq:V1V2}
\end{equation}
the vector $\vec{V_1}$ and $\vec{V_2}$ are not orthogonal unless $a=b$, i.e. $A$ and $B$ have the same variance.

\subsection{A particular case: $A$ and $B$ have the same variance\label{sec:particular_case}}
Let us define $\sigma_N^2=\sigma_A^2=\sigma_B^2=n^2$, i.e. $n=a=b$. In this case
$$
\mathbb{E}\left[\vec{V}_1\cdot\vec{V}_2\right]=\frac{n^2}{4}-\frac{n^2}{4}=0,
$$
$\vec{V}_1, \vec{V}_2$ are orthogonal which means that 
their squared modulus are 2 independent $\chi^2$ rv:
$$
||\vec{V}_1||^2=v_1^2\chi_{v1}^2 \quad \textrm{and} \quad ||\vec{V}_2||^2=v_2^2\chi_{v2}^2
$$
Thanks to \cite[Appendix A]{vernotte2019}, we know that this $\chi^2$ rv difference is a V$\Gamma$ rv with a Probability Density Function (PDF), introduced by \cite{madan1990}:
\begin{equation}
p(x)=\frac{\gamma^{2\lambda}|x-\mu|^{\lambda-1/2}K_{\lambda-1/2}\left(\alpha|x-\mu|\right)}{\sqrt{\pi}\Gamma(\lambda)(2\alpha)^{\lambda-1/2}}e^{\beta(x-\mu)}
\label{eq:pdf_VG}
\end{equation}
\noindent
where $\gamma=\sqrt{\alpha^2-\beta^2}$, $\Gamma(\lambda)$ is the gamma function, $K_w(z)$ is a hyperbolic Bessel function of the second kind ($w\in\mathbb{R}$ and $z\in\mathbb{C}$) and with the following parameters:
\begin{equation}
\mu=0, \quad \alpha=\frac{v_1^2+v_2^2}{4v_1^2 v_2^2}, \quad \beta=\frac{v_1^2-v_2^2}{4v_1^2 v_2^2}, \quad \lambda=\frac{1}{2}\label{eq:particular_case}
\end{equation}
where $\lambda$ is the number of dof divided by 2. In this particular case, since $a^2=b^2=n^2$, $v_1^2$ and $v_2^2$ becomes
$$
v_1^2=\mathbb{E}\left[||\vec{V}_1||^2\right]=\frac{n^2}{2}+c^2 \quad \textrm{and} \quad v_2^2=\mathbb{E}\left[||\vec{V}_2||^2\right]=\frac{n^2}{2},
$$
and we obtain
$$
\alpha=\frac{n^2+c^2}{n^2(2n^2+c^2)} \quad \textrm{and} \quad \beta=\frac{c^2}{n^2(2n^2+c^2)}
$$

\subsection{General case}
If $\sigma_A^2\neq\sigma_B^2$, $\vec{V}_1$ and $\vec{V}_2$ are no longer orthogonal and therefore they are 2 correlated rv. We have then to search another set of basis vectors which are orthogonal. For this purpose, let us use the Gram-Schmidt process.

	\subsubsection{Gram-Schmidt orthogonalization}

Let us keep $\vec{V}_1$ unchanged. Let $\vec{V}_{2P}$ be the projection of $\vec{V}_2$ onto $\vec{V}_1$. 
Denoting $\theta$ the angle\footnote{In the same way as the orthogonality between 2 vectors is defined by the null mathematical expectation of their scalar product, the angles as well as the other relationships between vectors must be taken into account as mathematical expectation since they are valid on average but not for only one particular realization of these vectors.} between $\vec{V}_1$ and $\vec{V}_2$, it comes
$$
\vec{V}_{2P}=v_2 \cos(\theta) \frac{\vec{V}_1}{v_1}.
$$
with
$$
\cos(\theta)=\frac{\mathbb{E}\left[\vec{V}_1\cdot\vec{V}_2\right]}{v_1v_2}
$$
and then
\begin{equation}
\vec{V}_{2P}=\frac{\mathbb{E}\left[\vec{V}_1\cdot\vec{V}_2\right]}{v_1^2}\vec{V}_1.\label{eq:proj}
\end{equation}

Therefore, we can build the vector $\vec{V}_{2N}$ which is the component of $\vec{V}_2$ orthogonal to $\vec{V}_1$:
$$
\vec{V}_{2N}=\vec{V}_2-\vec{V}_{2P}=\vec{V}_2-\frac{\mathbb{E}\left[\vec{V}_1\cdot\vec{V}_2\right]}{v_1^2}\vec{V}_1.
$$
Using Eq. (\ref{eq:v12v22}) and (\ref{eq:V1V2}) yields
\begin{equation*}
\begin{aligned}
\vec{V}_{2N}&=\left(\begin{array}{c}
a/2\\
-b/2\\
0
\end{array}\right)-\frac{a^2-b^2}{a^2+b^2+4c^2}\left(\begin{array}{c}
a/2\\
b/2\\
c
\end{array}\right)\\
&=\left(\begin{array}{c}
\displaystyle\frac{a(b^2+2c^2)}{a^2+b^2+4c^2}\\
\displaystyle-\frac{b(a^2+2c^2)}{a^2+b^2+4c^2}\\
\displaystyle-\frac{c(a^2-b^2)}{a^2+b^2+4c^2}
\end{array}\right)=\left(\begin{array}{c}
v_{2nA}\\
v_{2nB}\\
v_{2nC}
\end{array}\right).
\end{aligned}
\end{equation*}

We have now to express the measurement vectors $\vec{A}+\vec{C}$ and $\vec{B}+\vec{C}$ as linear combinations of the new basis of orthogonal vectors $\vec{V}_1$ and $\vec{V}_{2N}$. In order to do this, we must project these 2 measurement vectors onto the 2 basis vectors in the same way that we have projected $\vec{V}_2$ onto $\vec{V}_1$ in Eq. (\ref{eq:proj}):
$$
\left\{\begin{array}{lcl}
\vec{A}+\vec{C}&=&k_{AC1}\vec{V}_1+k_{AC2n}\vec{V}_{2N}\\
\vec{B}+\vec{C}&=&k_{BC1}\vec{V}_1+k_{BC2n}\vec{V}_{2N}
\end{array}\right.
$$
with
$$
\begin{array}{ll}
\displaystyle k_{AC1}=\frac{\mathbb{E}\left[\left(\vec{A}+\vec{C}\right)\cdot \vec{V}_1\right]}{\mathbb{E}\left[||\vec{V}_1||^2\right]}&\displaystyle k_{AC2n}=\frac{\mathbb{E}\left[\left(\vec{A}+\vec{C}\right)\cdot \vec{V}_{2N}\right]}{\mathbb{E}\left[||\vec{V}_{2N}||^2\right]}\\
\displaystyle k_{BC1}=\frac{\mathbb{E}\left[\left(\vec{B}+\vec{C}\right)\cdot \vec{V}_1\right]}{\mathbb{E}\left[||\vec{V}_1||^2\right]}&\displaystyle k_{BC2n}=\frac{\mathbb{E}\left[\left(\vec{B}+\vec{C}\right)\cdot \vec{V}_{2N}\right]}{\mathbb{E}\left[||\vec{V}_{2N}||^2\right]}.
\end{array}
$$
Therefore, $Z=(\vec{A}+\vec{C})\cdot(\vec{B}+\vec{C})$ may be written as
\begin{eqnarray}
Z&=&k_{AC1}k_{BC1}||\vec{V}_1||^2+k_{AC2n}k_{BC2n}||\vec{V}_{2N}||^2\nonumber\\
&&+\left(k_{AC1}k_{BC2n}+k_{AC2n}k_{BC1}\right)\vec{V}_1\cdot\vec{V}_{2N}\nonumber\\
&=&k_{AC1}k_{BC1}\dot{\chi}^2+k_{AC2n}k_{BC2n}\ddot{\chi}^2\nonumber\\
&&+\left(k_{AC1}k_{BC2n}+k_{AC2n}k_{BC1}\right)\textrm{V}\Gamma\label{eq:raw_quadform}
\end{eqnarray}
where $\dot{\chi}^2$ and $\ddot{\chi}^2$ are independent. Thus, this relationship involves the difference of 2 $\chi^2$ rv (it can be proved that $k_{AC2n}k_{BC2n}<0$), which is well known \cite{vernotte2019,sorin1968}, plus a V$\Gamma$ rv, which makes the problem more complex. In order to simplify this problem, we should find a representation of Eq. (\ref{eq:raw_quadform}) in which the cross term is identically null.

\subsubsection{Normalization and rotation of the basis vectors\label{sec:normalization}}
Let $(\vec{V}'_1,\vec{V}'_2)$ be the normalized equivalent of the basis $(\vec{V}_1,\vec{V}_{2N})$:
$$
\vec{V}'_1=\frac{\vec{V}_1}{\mathbb{E}\left[||\vec{V}_1||\right]} \quad \textrm{and} \quad \vec{V}'_2=\frac{\vec{V}_{2N}}{\mathbb{E}\left[||\vec{V}_{2N}||\right]}.
$$
With this new basis, Eq. (\ref{eq:raw_quadform}) may be rewritten as
\begin{eqnarray}
Z&=&k_{AC1}k_{BC1}\mathbb{E}\left[||\vec{V}_1||\right]^2||\vec{V}'_1||^2\nonumber\\
&&+k_{AC2n}k_{BC2n}\mathbb{E}\left[||\vec{V}_{2N}||\right]^2||\vec{V}'_2||^2\nonumber\\
&&+k_{AC1}k_{BC2n}\mathbb{E}\left[||\vec{V}_1||\right]\mathbb{E}\left[||\vec{V}_{2N}||\right]\vec{V}'_1\cdot\vec{V}'_2\nonumber\\
&&+k_{AC2n}k_{BC1}\mathbb{E}\left[||\vec{V}_1||\right]\mathbb{E}\left[||\vec{V}_{2N}||\right]\vec{V}'_1\cdot\vec{V}'_2\nonumber\\
&=&k'_{11}||\vec{V}'_1||^2-k'_{22}||\vec{V}'_2||^2+k'_{12}\vec{V}'_1\cdot\vec{V}'_2\label{eq:simple_quadform}
\end{eqnarray}
with
$$
\left\{\begin{array}{lcl}
k'_{11}&=&k_{AC1}k_{BC1}\mathbb{E}\left[||\vec{V}_1||\right]^2\\
k'_{22}&=&-k_{AC2n}k_{BC2n}\mathbb{E}\left[||\vec{V}_{2N}||\right]^2\\
k'_{12}&=&k_{AC1}k_{BC2n}\mathbb{E}\left[||\vec{V}_1||\right]\mathbb{E}\left[||\vec{V}_{2N}||\right]\\
&&+k_{AC2n}k_{BC1}\mathbb{E}\left[||\vec{V}_1||\right]\mathbb{E}\left[||\vec{V}_{2N}||\right].
\end{array}\right.
$$

We can then consider Eq. (\ref{eq:simple_quadform}) as the expression of a quadratic form $Q$ which associate a scalar $w_0$ to any vector $\vec{W}=w_1\vec{V}'_1+w_{2n}\vec{V}'_2$. Such a quadratic form may be described as
\begin{equation}
w_0=\vec{w}^T[Q]\vec{w} \quad \textrm{with} \quad [Q]=\left(\begin{array}{cc}
k'_{11} & k'_{12}/2\\
k'_{12}/2 & -k'_{22}
\end{array}\right).
\label{eq:nodiag_matrix}
\end{equation}
The simplification of our problem relies then in a rotation of the basis vectors in such a way that the quadratic form matrix $[Q]$ is diagonal. The eigenvalues of $[Q]$ are given by
$$
\ell_1=\frac{k'_{11}-k'_{22}-\sqrt{\Delta}}{2} \quad \textrm{and} \quad \ell_2=\frac{k'_{11}-k'_{22}+\sqrt{\Delta}}{2}.
$$
with $\Delta=\left(k'_{11}+k'_{22}\right)^2+k'^{\,2}_{12}$. Thanks to this rotation of the basis vectors, Eq. (\ref{eq:raw_quadform}) and (\ref{eq:simple_quadform}) become
$$
Z=\ell_1\dot{\chi}^2+\ell_2\ddot{\chi}^2.
$$
As already stated in {\S} \ref{sec:particular_case}, $Z$ is a V$\Gamma$ rv with the following PDF parameters:
\begin{equation}
\mu=0, \quad \alpha=\frac{\ell_1^2+\ell_2^2}{4\ell_1^2 \ell_2^2}, \quad \beta=\frac{\ell_1^2-\ell_2^2}{4\ell_1^2 \ell_2^2}, \quad \lambda=\frac{1}{2}. \label{eq:general}
\end{equation}

\subsection{Generalization to larger degrees of freedom}
In the case of $2m$ dof, i.e. real part + imaginary part multiplied by $m$ averaged uncorrelated spectra, the only change to apply concerns the parameter $\lambda$ in Eq. (\ref{eq:particular_case}) and (\ref{eq:general}) which becomes $\lambda=m$.\\
According to \cite[Eq. 12 p.80]{watson1922} we have the following relation:
\begin{equation}
    K_{n+\frac{1}{2}}(z)=\left(\frac{\pi}{2z}\right)^{\frac{1}{2}}e^{-z}\sum_{r=0}^{n}\frac{(n+r)!}{r!(n-r)!(2z)^r}
    \label{eq:BesselK}
\end{equation}
\noindent
with $n\in\mathbb{N}$ and $z\in\mathbb{C}$. Moreover $m\in\mathbb{N}^*$ which leads to the relation $n=m-1$. Therefore let us expand Eq. (\ref{eq:pdf_VG}) using Eq. (\ref{eq:BesselK}):

\begin{equation}
    p(x)=\frac{\kappa(\alpha,\beta)^m\epsilon(x,\mu,\alpha,m)}{\Gamma(m)}e^{-\alpha|x-\mu|+\beta(x-\mu)}
    \label{eq:p(x)}
\end{equation}
\noindent
with the following parameters:
$$
\kappa(\alpha,\beta)=\frac{\alpha^2-\beta^2}{2\alpha} \quad \Gamma(m)=(m-1)!
$$
$$
\epsilon(x,\mu,\alpha,m)=\sum_{r=0}^{m-1}\frac{(m+r-1)!|x-\mu|^{m-r-1}}{r!(m-r-1)!(2\alpha)^r}
$$

\subsection{Validation of the theoretical probability laws by Monte Carlo simulations}
\subsubsection{Algorithm description\label{sec:algo}}

According to {\S} \ref{sec:normalization} the probability density of $Z$, equal to the difference of two independent $\chi^2$ rv, can now be calculated using the function $p(x)$ of the Eq. (\ref{eq:p(x)}) by assigning the values to the parameters in Eq. (\ref{eq:particular_case}) and (\ref{eq:general}). In order to perform this comparison we use two algorithms, one for Monte Carlo (MC) simulation and the other one for computing Eq. (\ref{eq:p(x)}).\\

\noindent
$-$ \textit{MC simulation algorithm}\\
\indent
The simulation algorithm follows these 6 steps
\begin{enumerate}[S1:]
\item Assignement of the 2 noise levels $\sigma_A^2$, $\sigma_B^2$, signal level $\sigma_C^2$ and the number of averaging spectra $m$.
\item Drawing of A, B, C, following a normal centered distribution with respectively $\sigma_A$, $\sigma_B$, $\sigma_C$ as standard deviation.
\item Computation of  $Z=(A+C)(B+C)$.
\item Repetition $2m$ times of the steps S2 to S3 and sum all $Z$ values.
\item Repetition $N=10^7$ times of the steps S2 to S4 of this sequence.
\item Drawing the histogram of $Z$.\\
\end{enumerate}

In all simulations, we chose a number of dof $\nu=2m$ in order to have a real and imaginary part in agreement with the experiment shown in Fig. (\ref{fig1}).\\

\noindent
$-$ \textit{Modeling algorithm}\\
The modeling algorithm follows also 6 steps:
\begin{enumerate}[S1:]
\item Assignement of the 2 noise levels $\sigma_A^2$, $\sigma_B^2$, signal level $\sigma_C^2$ and the number of averaging spectra $m$.
\item \textbf{Independent basis}
\begin{itemize}
    \item Computation of coefficients $v_1^2$, $v_2^2$ according to Eq. (\ref{eq:v12v22})
    \item if $\sigma_A^2=\sigma_B^2$ go to step S5 else perform steps S3 and S4
\end{itemize}
\item \textbf{Orthogonalization of the basis}
\begin{itemize}
    \item Computation of coefficients $k_{AC1}$, $k_{AC2n}$, $k_{BC1}$, $k_{BC2n}$ to determine the new basis according to Eq. (\ref{eq:raw_quadform})
    \item Normalization of the basis by determing coefficients $k'_{11}$, $k'_{22}$, $k'_{12}$ according to Eq. (\ref{eq:simple_quadform})
\end{itemize}
\item \textbf{Vector rotation}
\begin{itemize}
    \item Diagonalization of the matrix $Q$ according to Eq. (\ref{eq:nodiag_matrix})
    \item Computation of its roots $l_1$ and $l_2$
\end{itemize}
\item Compute the coefficients $\alpha$, $\beta$, and $\lambda=m$ according to Eq. (\ref{eq:particular_case}) and (\ref{eq:general}).
\item Plotting the probability density with Eq. (\ref{eq:p(x)}).\\
\end{enumerate}

\subsubsection{When can the instrument noises be assumed to be ``about the same''?}

Although the problem is quite simple when the instrument noises $\sigma_A^2$ and $\sigma_B^2$ are the same (see {\S} \ref{sec:particular_case}), it becomes more complex when $\sigma_A^2\neq\sigma_B^2$. The question is then how far can we assume that $
\sigma_A^2\approx\sigma_B^2$ and then use the particular case formalism of {\S} \ref{sec:particular_case}? In order to answer this question, we use Monte-Carlo simulations which were performed according to {\S} \ref{sec:algo}.

\begin{figure}
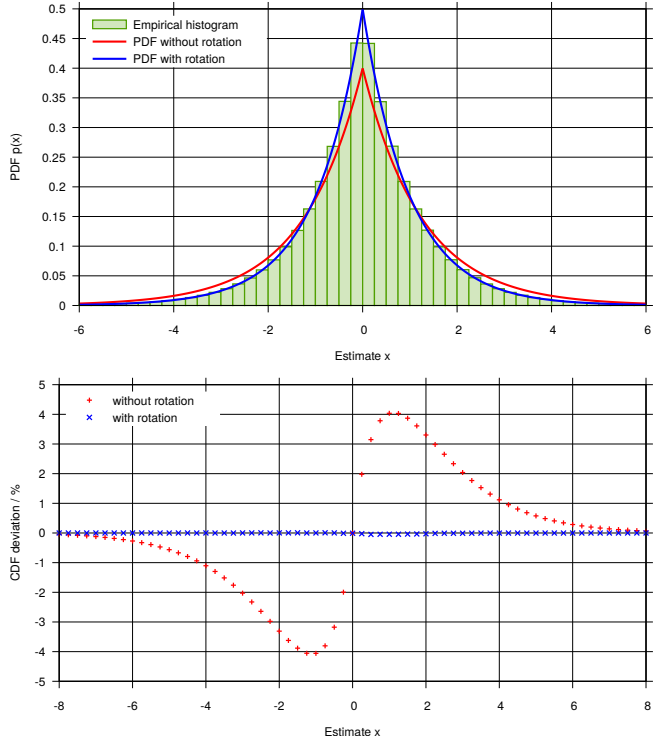

\centering
\includegraphics[height=5cm]{histo_crospec3.pdf}\\
\includegraphics[height=5cm]{cdf_dev3.pdf}
\caption{Comparison of the empirical and theoretical PDF (above) with and without rotation of the basis vectors. The deviations between the empirical and the theoretical CDF are given in the bottom plot. The variances are: $\sigma_C^2=0, \sigma_A^2=2, \sigma_B^2=1/2$ and there are 2 dof.\label{fig:comp_emp_theo}}
\end{figure}

Afterwards we perform a histogram of these realizations and compare it first with the PDF obtained from the model without rotation, i.e. by using the V$\Gamma$ parameters of Eq. (\ref{eq:particular_case}), and next with the PDF obtained from the model with rotation, i.e. by using the V$\Gamma$ parameters of Eq. (\ref{eq:general}). Fig. \ref{fig:comp_emp_theo} shows an example of such a comparison. In this case ($\sigma_A^2=2, \sigma_B^2=1/2, \sigma_C^2=0$), the PDF of the model with rotation is in perfect agreement with the histogram whereas there are large discrepancies with the PDF of the model without rotation. We have thus a first result: the model without rotation should not be used when the ratio $\sigma_A^2/\sigma_B^2\geq 4$.

To improve the efficiency of the test, we compute the theoretical quantiles by using the model without rotation and then deduce from them the theoretical confidence intervals which are often used (68 \%, 90 \%, 95 \% and 99 \%). These quantiles and intervals are compared to the ones obtained from the simulation histogram. In one example of Table \ref{tab:all}, which corresponds to the case plotted in Fig. \ref{fig:comp_emp_theo}, the confidence intervals are strongly overestimated. For instance, the expected 68 \% confidence interval is significantly too large since it encompasses an interval of 76 \%. Similarly, the expected \mbox{90 \%} interval is actually a 94 \% interval. This reinforces our decision of using the model with rotation for a noise variance ratio $\geq 4$.

\begin{table}[t!]
\begin{center}
    \caption{Comparison of the expected quantiles and intervals\label{tab:all}}
\begin{tabular}{||r||rrrr|rr||}
\hline
\hline
\multirow{4}{1.6cm}{\textbf{Expected probabilities (\%)}}&\multicolumn{6}{c||}{\textbf{True probabilities (\%)}}\\
\cline{2-7}
&\multicolumn{4}{c|}{Degrees of freedom: 2}&\multicolumn{2}{c||}{dof: 8}\\
\cline{2-7}
&\multicolumn{4}{c|}{$\sigma_C^2=$0}&0&0.5\\
\cline{2-7}
& $\sigma_B^2=$2 & 1 & 2/3 & 1/2  & 1 & 1\\
\hline
\hline
\multicolumn{1}{||l||}{Quantiles}&&&&&&\\
0.5 &  0.50 &  0.39 &  0.25 &  0.16 &  0.35 &  0.39 \\
2.5 &  2.50 &  2.10 &  1.58 &  1.19 &  1.98 &  2.13 \\
5.0 &  5.00 &  4.36 &  3.51 &  2.82 &  4.18 &  4.44 \\
16.0 & 16.00 & 14.95 & 13.43 & 12.05 & 14.68 & 15.18 \\
50.0 & 50.00 & 50.01 & 50.00 & 50.00 & 50.00 & 49.99 \\
84.0 & 84.00 & 85.07 & 86.59 & 87.96 & 85.32 & 84.58 \\
95.0 & 95.00 & 95.65 & 96.50 & 97.19 & 95.82 & 95.37 \\
97.5 & 97.50 & 97.91 & 98.43 & 98.82 & 98.02 & 97.73 \\
99.5 & 99.50 & 99.62 & 99.75 & 99.84 & 99.65 & 99.57 \\
\hline
\hline
\multicolumn{1}{||l||}{Intervals}&&&&&&\\
68.0 & 68.00 & 70.12 & 73.16 & 75.91 & 70.64 & 69.41 \\
90.0 & 90.00 & 91.29 & 92.98 & 94.37 & 91.64 & 90.93 \\
95.0 & 95.00 & 95.82 & 96.84 & 97.63 & 96.04 & 95.60 \\
99.0 & 99.00 & 99.23 & 99.50 & 99.68 & 99.30 & 99.18 \\
\hline
\hline
\end{tabular}
\end{center}
{The expected quantiles (above) and intervals (below) are computed by using the parameters from Eq. (\ref{eq:particular_case}) with empirical probabilities. For all realizations $\sigma_A^2=2$.}
\end{table}

We use these 2 approaches, i.e. PDF curve as well as confidence intervals, for many different parameter sets (see Table \ref{tab:all}). In any case, the agreement between the model with rotation and the Monte-Carlo simulation histograms were perfect, since the residual deviations can be largely assumed to be due to the finite sample number of the simulation (less than 0.05 \% of the CDF). However, this test is very interesting for the model without rotation since it allows us to answer to the question which is the title of this section: when can the instrument noises be assumed to be ``about the same''? Table \ref{tab:all} is very useful in this connection. In a first step, let us study the case where the number of dof is 2 and there is no signal since it is the case which is the most sensitive to the difference between the noise levels. We can see on this table that the model without rotation is perfect when the 2 noise levels are equal ($\sigma_B^2=2$), fair when the ratio of the noise levels is equal to 2 ($\sigma_B^2=1$), at the limit of acceptance when the ratio is 3 but not suitable for a ratio $\geq 4$. The other columns of Table \ref{tab:all}, obtained with 8 dof and with $\sigma_C^2=\sigma_A^2/4$, confirm that the model without rotation is acceptable when the ratio of the noise variances is equal to 2.

Then we recommend to use the vector rotation process if the ratio of the noise variance greater than 2.

\section{Inverse Problem\label{sec3}}
\subsection{Principle of the method}

The bayesian statistician has to solve the inverse problem in order to define a confidence interval for the true variance $\sigma_C^2$, given a set of measurements and a priori information. Thereby the cross-spectrum measurement $Z$ is now fixed as well as the instrument noise levels $\sigma_A^2$ and $\sigma_B^2$, whereas the signal true variance $\sigma_C^2$ appears as a random variable. According to the Bayes theorem the a posteriori density of an unknown true value $\theta$ given th measurements, here the cross-spectrum $Z$, is

\begin{equation}
\left\{\begin{array}{l}
p(\theta|Z) \propto p(Z|\theta)\cdot\pi(\theta)\\
\int_0^\infty p(\theta|Z) d\theta=1
\end{array}\right.
\end{equation}
\noindent
where $\pi(\theta)$ is the a priori density, named prior and $p(Z|\theta)$ is the PDF which corresponds to Eq. (\ref{eq:p(x)}) determined in the direct problem. It remains to determine the prior $\pi(\theta)$ (i.e. the PDF before any measurement) to compute the a posteriori density.

One of the main issue of Bayesian analysis concerns the choice of this prior. We have no a priori knowledge about the behavior of the parameter $\theta$. A total ignorance of knowledge leads to a prior equal to $\theta^{-1}$ which means all order of magnitudes have the same probability. The choice of $\theta$ is subject to discussion and the reader should refer to \cite[Appendix B]{mchugh96b}.

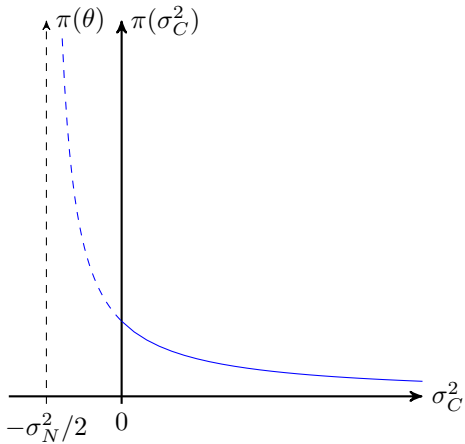
\begin{figure}[H]
\begin{center}
\begin{tikzpicture}
\begin{scope}[scale=1]
\draw[thick,->,>=stealth'] (-0.5,0)--(5,0);
\draw (5,0) node[right]{$\sigma_C^2$};
\draw[dashed,->,>=stealth'] (0,-.1)--(0,5);
\draw (0,-0.1) node [below]{$-\sigma_N^2/2$};
\draw (0,5) node[right]{$\pi(\theta)$};
\draw[thick,->,>=stealth'] (1,-.1)--(1,5);
\draw (1,-0.1) node [below]{$0$};
\draw (1,5) node[right]{$\pi(\sigma_C^2)$};
\draw[blue, dashed, domain=0.21:1] plot ({\x}, {1/\x)});
\draw[blue, domain=1:5] plot ({\x}, {1/\x)});
\end{scope}
\end{tikzpicture}
\caption{Prior of the sum of the noise $\sigma_N^2$ and signal $\sigma_C^2$ levels for the case when there is no averaging spectra (i.e. $\nu=2$).\label{fig:prior}}
\end{center}
\end{figure}

The quantity that can be actually measured is the sum of the signal and the measurement noise. Hence the prior should be accordingly given as a function of this sum. In other words, it is not possible to have any information on a signal with a level much smaller than the measurement noise. Hence choosing a prior function of $\sigma_N^2+\sigma_C^2$ ensures that the corresponding magnitude order of $\sigma_C^2$ do not dominate the a posteriori probability distribution. The measurement noise level decreases as $m^{-1}$, according to \cite[Eq. 11]{rubiola2010}, when averaging over different spectra realizations $m$. So it should depend to the number of dof $\nu=2m$ (i.e. taking in account the real and imaginary part). From this considerations, we chose the following prior according to Fig. \ref{fig:prior}:

\begin{equation}
\pi(\theta) = \frac{1}{\theta} = \frac{1}{\sigma_N^2/\nu+\sigma_C^2},\label{eq:prior}
\end{equation}
\noindent
where $\sigma_N^2=(\sigma_A^2+\sigma_B^2)/2$ is the known, ``not random'' averaged noise level. Thus small level of $\sigma_C^2$ are distributed roughly uniformly on a linear scale and large values are distributed with equal probability for equal logarithmic intervals. 

\subsection{Check of the posterior probability density function}
According to Eq. (\ref{eq:p(x)}), for 2 dof or $m=1$ spectrum average and the particular case $\sigma_A^2=\sigma_B^2=\sigma_N^2$, we know that
$$
p\left(Z|\sigma_c^2\right)=\frac{e^{Z/\sigma_N^2}}{2(\sigma_N^2+\sigma_C^2)}.
$$
Therefore, the posterior PDF of the cross-spectrum estimator is
\begin{equation}
\left\{\begin{array}{lll}
p\left(\sigma_C^2|Z\right)\propto \displaystyle\frac{e^{Z/\sigma_N^2}}{2(\sigma_N^2+\sigma_C^2)(\sigma_N^2+2\sigma_C^2)} &\textrm{if}& Z\leq 0\\
&&\\
p\left(\sigma_C^2|Z\right) \propto \displaystyle\frac{e^{-Z/(\sigma_N^2+2\sigma_C^2)}}{2(\sigma_N^2+\sigma_C^2)(\sigma_N^2+2\sigma_C^2)} &\textrm{if}& Z\geq 0.
\end{array}\right.\label{eq:post}
\end{equation}

We have checked this posterior PDF by using the inverse problem Monte-Carlo algorithm we already used in \cite[{\S} IV.B.1)]{vernotte2019} and \cite[{\S} IV.A.]{lantz2019}. The principle is the following:

\noindent
\begin{enumerate}[S1:]
\item Select a target estimate $Z=Z_0$.
\item Draw at random the signal level $\sigma_C^2$ according to
$$
\sigma_C^2=10^{\left[\eta+u_{[0,1]}(e_{max}-\eta)\right]}-\frac{\sigma_N^2}{2}
$$
where $\eta=\log_{10}(\sigma_N^2/2)$ and $u_{[0,1]}$ is a pseudo-random function which is uniform within $[0,1]$. This draw ensures the parameter follows the prior of Eq. (\ref{eq:prior}) up to $10^{e_{max}}$. We have chosen $e_{max}=4$ which is in accordance with Fig. \ref{fig:prior}
\item Draw at random (Gaussian) the noise and signal estimates $A,B,C$ and compute the measurements $X, Y$ according to Eq. (\ref{eq8}).
\item Compute the estimate $Z$.
\item Compare the estimate $Z$ with the target $Z_0$: if $Z= Z_0 \pm p $, store the current $\sigma_C^2$ value as it is able to generate an estimate equal to the target; otherwise throw this $\sigma_C^2$ value. We have chosen $p=(Z_0+\sigma_N^2/2)/50$ when $Z_0>0$ and $p=\sigma_N^2/100$ when $Z_0\leq 0$.
\item Go to step 2.
\item Stop when a set of 10\,000 $\sigma_C^2$ values is reached.
\end{enumerate}

\begin{figure}[H]
\includegraphics[width=\linewidth]{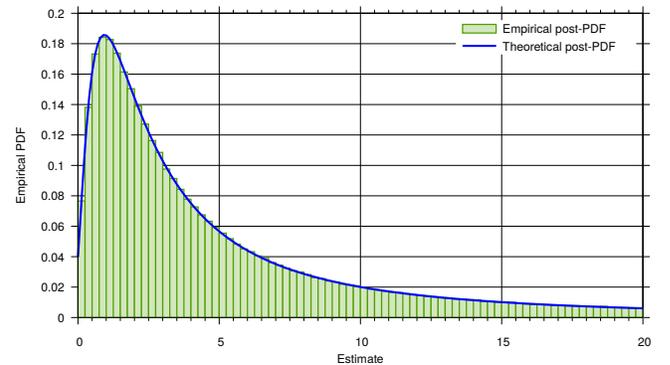}
\caption{Comparison of the empirical and theoretical posterior PDF for a noise level $\sigma_N^2=1$ a.u and a target estimate $Z_0=1$ a.u.\label{fig:CSpdfnorm}}
\end{figure}

\begin{table}
\begin{center}
\caption{Comparison of the quantiles 95 \% obtained by Monte-Carlo simulation and by the posterior CDF\label{tab:quantiles}}
\begin{tabular}{c|rr|c}
\multicolumn{1}{c|}{Target $Z_0$ (a.u.)} & \multicolumn{2}{c|}{95 \% bound}  & \multicolumn{1}{c}{True prob. (\%)}\\
\multicolumn{1}{c|}{}& Emp & Theo & \multicolumn{1}{c}{}\\
\hline
-1.00 & 14.04 & 13.65 & 94.90 \\
 0.00 & 15.11 & 13.65 & 94.53 \\
 0.10 & 14.52 & 14.27 & 94.91 \\
 0.20 & 14.98 & 14.90 & 94.96 \\
 0.32 & 15.87 & 15.66 & 94.94 \\
 0.50 & 17.37 & 16.90 & 94.87 \\
 1.00 & 20.40 & 20.14 & 94.93 \\
 2.00 & 27.61 & 28.65 & 95.18 \\
 3.16 & 38.19 & 39.08 & 95.08 \\
 5.00 & 57.15 & 56.55 & 94.91 \\
10.00 & 109.66 & 104.82 & 94.78
\end{tabular}
\end{center}
{The quantiles 95 \% are computed for a noise level $\sigma_N^2=1$ a.u. The theoretical bounds (denoted ``Theo") are obtained by numerical integration and then correspond to the true probabilities (denoted ``True prob.").}
\end{table}

It must be noticed that such an algorithm is obviously not able to justify the choice of the prior since this prior is included in the algorithm. It will only ensure that no mistake has been done in the expression of the posterior PDF.

Fig. \ref{fig:CSpdfnorm} shows the comparison of the posterior PDF computed according to Eq. (\ref{eq:post}) (blue curve) and the histogram obtained thanks to the inverse problem Monte-Carlo algorithm (green boxes) with a noise level $\sigma_N^2=1$ a.u and a target estimate $Z_0=1$ a.u. We can verify that the agreement is excellent.

Table \ref{tab:quantiles} compares the $95 \%$ quantiles obtained by the inverse problem Monte-Carlo algorithm (denoted ``Emp'' for empirical) and by the integration of the posterior PDF (denoted ``Theo'' for theoretical), i.e. the posterior CDF, for different values of target and for a noise level $\sigma_N^2=1$. Here also the agreement is very good whether for the $95 \%$ bounds or for the true probabilities of the theoretical bounds. Moreover, the fluctuations of the empirical bounds proves that the slight differences between empirical and theoretical values are due to the fluctuations of the empirical bounds because of the limited number of realizations ($10\,000$) of the inverse problem Monte-Carlo algorithm.

\subsection{KLT method}

The KLT method stands for ``Karhunen-Lo\`{e}ve Tranform" and was developed in our previous paper \cite{lantz2019}. In that paper, KLT has proved to be as efficient as well as rigorous method, making the most of the property of ``sufficient statistics". However the difference with \cite{lantz2019} is that we don't have the ``\textit{sufficient statistics}" property (see \cite{saporta1990}). It means that KLT method will not give the same result as the cross-sprectrum method whereas it should have in the case of ``\textit{sufficient statistics}". First let us remind the theory. Then in a second time, we will explain what can bring the KLT method in addition to the cross-specrum one.\\

\subsubsection{A Posteriori distribution}

The KLT method relies on the use of $X$, and $Y$ measurements according to Eq. (\ref{eq2}), which are Gaussian rv instead of the product of $AB$, $AC$, $BC$ and $C^2$, which are linear combination of Bessel of the second kind functions and $\chi^2$ random variables. The main advantage of this approach lays in the property of the Gaussian rv which remain Gaussian when they are linearly combined.
However, these measurements are not independent. That is why we aim to determine two linear combinations of these rv that are independent one of each other. Hence we define the covariance matrix between $X$ and $Y$ given by

\begin{equation}
M=\begin{pmatrix}
\sigma_A^2+\sigma_C^2 & \sigma_C^2\\
\sigma_C^2 & \sigma_B^2+\sigma_C^2
\end{pmatrix}.
\end{equation}

The KLT consists in using the rv corresponding to the diagonalization of this matrix. In order to simplify the equations we study solely the case where $\sigma_A^2=\sigma_B^2=\sigma_N^2$. The eigenvalues of $M$ are
%
%

\begin{equation}
    \begin{aligned}
        \lambda_1 &= \sigma_N^2+2\sigma_C^2\\
        \lambda_2 &= \sigma_N^2
    \end{aligned}
\end{equation}

with the following normalized eigenvectors,

\begin{equation}
\begin{array}{lcl}
V_1=\frac{1}{\sqrt{2}}\begin{pmatrix}
1\\
1
\end{pmatrix}
& & V_2=\frac{1}{\sqrt{2}}\begin{pmatrix}
1\\
-1
\end{pmatrix}
\end{array}
\end{equation}

The likelihood function is then given by

\begin{equation}
p_{KLT}(Z|\sigma_C^2)=\prod_{i=1}^{2}\frac{1}{\lambda_i^{\nu/2}}\exp\left(-\frac{\sum_{j=1}^\nu w_{i,j}^2}{2\lambda_i}\right)
\end{equation}

The numerator of the exponential argument is then the only term that depends on the actual measurements:

\begin{equation}
w_{i,j}^2=V_{i,1}^2 X_j^2 + V_{i,2}^2 Y_j^2 + 2V_{i,1}V_{i,2}Z_j
\end{equation}
\noindent
where $||V_i||^2=\sum_j V_{i,j}^2$. So the KLT method involve the spectral density $X^2$, $Y^2$ in addition to the cross-spectrum.

Keeping the same prior defined in Eq. (\ref{eq:prior}) we have the following a posteriori density,

\begin{equation}
\left\{\begin{aligned}
& p_{KLT}(\sigma_C^2|Z)\propto\frac{1}{\sigma_N^2/2+\sigma_C^2}\cdot p_{KLT}(Z|\sigma_C^2)\\
& \int_{\mathbb{R}}p_{KLT}(\sigma_C^2|Z)d\sigma_C^2=1.
\end{aligned}\right.
\end{equation}

\subsubsection{Validation of the method by Monte Carlo simulation}

In order to validate the KLT method, we have compared its results to Monte Carlo simulations. The algorithm is as follows:
\begin{enumerate}[S1:]
\item Select a noise level $\sigma_N^2=\sigma_A^2=\sigma_B^2$, a target $Z=Z_0$ and a combination $X=X_0$, $Y=Y_0=Z_0/X_0$ for all the dof.
\item Draw at random the signal level $\sigma_C^2$ according to
$$
\sigma_C^2=10^{\left[\eta+u_{[0,1]}(e_{max}-\eta)\right]}-\frac{\sigma_N^2}{2}
$$
where $\eta=\log_{10}(\sigma_N^2/2)$ and $u_{[0,1]}$ is a pseudo-random function which is uniform within $[0,1]$. This draw ensures the parameter follows the prior of Eq. (\ref{eq:prior}) up to $10^{e_{max}}$. We have chosen $e_{max}=4$.
\item Draw at random (Gaussian) the noise and signal estimates $A,B,C$ and compute the measurements $X, Y$ according to Eq. (\ref{eq8}).
\item Compute the estimates $X$ and $Y$.
\item Compare the estimates $X$, $Y$ with the targets $X_0$, $Y_0$ for all the dof: if $X= X_0 \pm p $, $Y= Y_0 \pm q $, store the current $\sigma_C^2$ value as it is able to generate an estimate equal to the target; otherwise throw this $\sigma_C^2$ value. We have chosen a precision $p$, $q$ of tenths of respectively $X_0$ and $Y_0$.
\item Go to step 2.
\item Stop when a set of $n$ $\sigma_C^2$ values is reached. The number of values $n$ depending on the computation time.\\
\end{enumerate}

\subsubsection{Results and discussion}

\begin{figure}
\includegraphics[width=\linewidth]{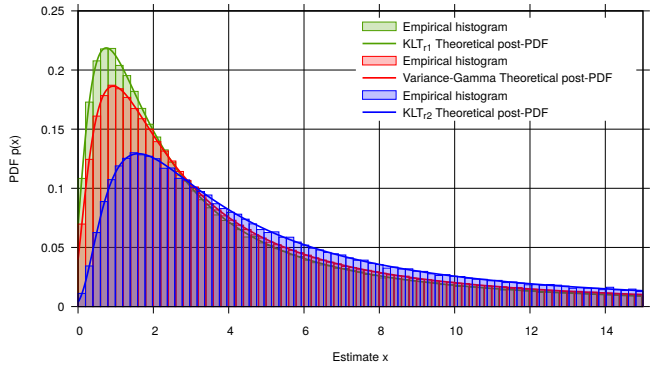}
\caption{Comparison of the empirical and theoretical posterior PDF for V$\Gamma$ and KLT methods with a noise level $\sigma_N^2=1$ a.u and a target estimate $Z_0=5$ a.u. KLT$_{r1}$ and KLT$_{r2}$ are the same method but differ by their combination of spectral density $X^2$ and $Y^2$ which are fixed, see Table \ref{tab:KLTS}, whereas they are rv for the V$\Gamma$ method.\label{fig:KLTS}}
\end{figure}

Fig. \ref{fig:KLTS} shows the comparison between the PDF of V$\Gamma$ method developed in $\S$ \ref{s:direct} and the PDF of KLT method for two different realizations. The theoretical post-PDF fits very well the empirical histogram for each method. The ``sufficient statistics" property being not valid, different combinations of the spectral density $X$ and $Y$ were tested and are given in Table \ref{tab:KLTS}. Indeed KLT$_{r1}$ and KLT$_{r2}$ realizations do not give the same PDF unlike the V$\Gamma$ method. KLT$_{r1}$ has then a peak which is higher than the V$\Gamma$ method whereas KLT$_{r2}$ has a smaller one. This is explained by a more stringent confidence interval for KLT$_{r1}$ than V$\Gamma$, and a less stringent for KLT$_{r2}$ as refered in Table \ref{tab:KLTS}. The 95 \% quantiles obtained with MC simulations are in a good agreement with the theoretical ones, especially for KLT$_{r1}$ and V$\Gamma$ methods. It is explained by the number of data which is not the same for all of these simulations. V$\Gamma$, KLT$_{r1}$ and KLT$_{r2}$ have respectively $1\,000\,000$, $500\,000$ and $245\,000$ data. V$\Gamma$ MC simulations takes only 2 minutes whereas it needs respectively 54 hours and 35 days using 17 cores, for KLT$_{r1}$ and KLT$_{r2}$. KLT$_{r1}$ is chosen to have the spectral density combination which leads to the most stringent confidence interval. Whereas KLT$_{r2}$ is chosen to be more defavourable than the general case V$\Gamma$ using only the knowledge of the cross-spectre measurement.

The KLT method can then have a slightly more stringent confidence interval than the cross-spectrum method using V$\Gamma$ for certain case. However it requires to have the knowledge of both spectral density of each channel. It then uses more information, the ``sufficient statistics'' property being not valid. So the KLT method is preferred when the spectral density are known.

\begin{table}
\begin{center}
    \caption{Comparison of the 95 \% quantiles obtained by Monte-Carlo simulation (empirical) and by the posterior CDF\label{tab:KLTS}}
\begin{tabular}{c|rrrr|rr}
\multicolumn{1}{c|}{Method} & \multicolumn{4}{c|}{Measurement} & \multicolumn{2}{c}{95 \% bound} \\
\multicolumn{1}{c|}{}& $X'$ & $X''$ & $Y'$ & $Y''$ & Emp & Theo \\
\hline
V$\Gamma$ & rv & rv & rv & rv & 56.4 & 56.6 \\
KLT$_{r1}$ & 1.6 & 1.6 & 1.6 & 1.6 & 48.4 & 48.3 \\
KLT$_{r2}$ & 4.0 & 0.6 & 2.5 & 1.0 & 82.3 & 80.8 \\
\end{tabular}
\end{center}
{The 95 \% quantiles are computed for a noise level $\sigma_N^2=1$ and a target estimate $Z_0=5$ a.u.}
\end{table}

\section{Conclusion\label{sec4}}

The method developed, V$\Gamma$, provides the Probability Density Function of the signal level studied when using the cross-spectrum method. It allows the determination of confidence intervals through numerical integration, where only the high bound has a physical meaning. It is especially relevant for one or several measurements of the cross-spectrum as the PDF will tend to a Gaussian distribution for many dof.

V$\Gamma$ is a rigorous method since it is the exact density solution of the cross-spectrum real part statistics, with no approximation. We shall notice that the noise level of each measurement instruments has to be known. If these noise level are the same except at a factor of 4 and higher, we can assume that all the theoretical part of orthogonalizing and the rotation of the basis is not necessary. This method works whatever the number of measurement spectra and noise level.

However using KLT method to compute the confidence interval is a more rigorous method because it uses the knowledge of the spectral density in addition to the cross-spectrum. That is why we recommend to use the KLT method which turns out to be a slightly better estimator than V$\Gamma$.


\section{Acknowledgement}

This work was partially funded by the ANR Programmes d'Investissement d'Avenir (PIA) Oscillator IMP (Project 11-EQPX-0033) and FIRST-TF (Project 10-LABX-0048).

\bibliography{Ref-local}

\end{document}